\documentclass[aps,prl,showpacs,preprint]{revtex4}
\usepackage{graphicx}

\input{epsf}

\begin{document}
\draft
\title{Creation of classical and quantum fluxons by a current dipole in a long
Josephson junction}
\author{Boris A. Malomed}
\address{Department of Interdisciplinary Studies, Faculty of Engineering,\\
Tel Aviv University \\
Tel Aviv 69978, Israel}
\author{Alexey V. Ustinov}
\address{Physikalisches Institut III, Universit\"at\\
Erlangen-N\"urnberg\\
D-91058, Erlangen, Germany}

\begin{abstract}
We study static and dynamical properties of fluxons in a long annular
Josephson junction (JJ) with a current injected at one point and collected
back at a close point. Uniformly distributed dc bias current with density
$\gamma$ is applied too. We demonstrate that, in the limit of the infinitely
small size of the current dipole, the critical value of $\gamma$, above
which static phase distributions do not exist, that was recently found (in
the Fraunhofer's analytical form) for the annular JJ with the length much
smaller than the Josephson penetration length, is valid irrespective of the
junction's length, including infinitely long JJs. In a long annular JJ, the
dipole generates free fluxon(s) if $\gamma$ exceeds the critical value. For
long JJs, we also find another critical value (in an analytical form too),
which is always slightly smaller than the Fraunhofer value, except for
points where the dipole strength is $2\pi N$ with integer $N$, and both
values vanish. The static phase configuration which yields the new critical
value is based on an unstable fluxon-antifluxon bound state, therefore it
will probably not manifest itself in the usual (classical) regime. However,
it provides for a dominating {\it instanton} configuration for tunnel birth
of a free fluxon, hence it is expected to determine a quantum-birth
threshold for fluxons at ultra-low temperatures. We also consider the
interaction of a free fluxon with the complex consisting of the current
dipole and antifluxon pinned by it. A condition for suppression of the net
interaction force, which makes the long JJ nearly homogeneous for the free
fluxon, is obtained in an analytical form. The analytical results are
compared with numerical simulations. The analysis presented in the paper is
relevant to the recently proposed new experimental technique of inserting
fluxons into annular Josephson junctions.
\end{abstract}


\date{\today}
\maketitle



\section{Introduction}

Long Josephson junctions (JJs) are well known to be a unique
physical system that allows one to experimentally study dynamics
of topological solitons in the form of fluxons (alias Josephson
vortices, each carrying a magnetic-flux quantum $\Phi _{0}$)
\cite{Barone,Ust-rev98}. For both experimental and theoretical
studies, the most convenient object is an {\em annular} (circular)
long JJ, in which the net number of initially trapped fluxons is
conserved, hence new solitons may only be created as
fluxon-antifluxon pairs \cite{Davidson:PRL85,Davidson:JAP86}.
Fluxon dynamics in the annular junctions manifests itself in the
clearest way, as it is not complicated by reflections from
boundaries. The interest to annular long JJs stems from the great
potential they offer for fundamental studies of dynamical
properties of solitons, such as, e.g., emission of the Cherenkov
radiation \cite{Goldobin-PRB98,Wallraff-PRL00}. A challenging
problem for the theory is quantum soliton dynamics in long JJs,
which has been recently observed in experiment at ultra-low
temperatures \cite{Nature-vortex}. On the other hand, annular JJs
offer applications in cryoelectronics, such as sources of highly
coherent microwave radiation \cite{Davidson:JAP86} and radiation
detectors \cite{radiat-detect}. Besides that, long annular JJs
have a potential for designing fluxon qubits \cite
{Wallraff00a,Kemp02b} and fluxon ratchets
\cite{Edik-PRE01,Carapella-PRB01}.

A key problem in experiments with annular JJs is the trapping of a
desired number of fluxons. The only previously known method which
made it possible to do that in a controllable way is rather
complicated, relying upon the use of a scanning electron (or
laser) microscope \cite{Ustinov-annular1}. Other studies relied on
fluxon trapping in the course of cooling the system below the
critical temperature $T_{{\rm c}}$ of the superconducting
electrodes
\cite{Davidson:PRL85,Davidson:JAP86,Goldobin-PRB98,Wallraff-PRL00,Vernik-annular1,others}.
However, the latter technique does not provide for reproducible
results, and requires heating of the junction to high
temperatures. Moreover, trapped fluxons generated in such a manner
get often pinned by Abrikosov vortices, which may be trapped in
the electrodes in the course of cooling below $T_{{\rm c}}$.

\begin{figure}[tbh]
\centering
\includegraphics[width=2.0in]{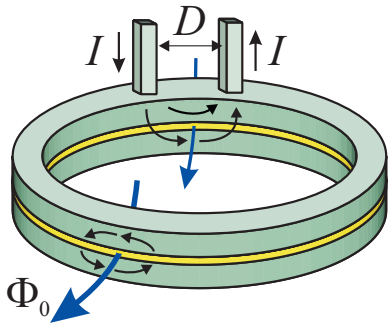}
\caption{A sketch of the annular Josephson junction with a current
dipole formed by local injection of current $I$ into one of the
superconducting electrodes.} \label{idea}
\end{figure}

In a recent work \cite{AVU}, a new method to insert fluxons into annular JJs
was proposed and demonstrated experimentally and numerically. It is based on
injecting a relatively large current $I$ into the junction locally; the
current flows from the injection point into the superconducting electrode
and also across the Josephson barrier, and is collected back at another
point of the same electrode, which is separated from the injection spot by a
small distance $D$. The schematic view of such JJ is given in
Fig.~\ref{idea}. In the following, we use the normalized notation
for the distance between
the current injectors, $d=D/\lambda _{J}$, and the current, $\varepsilon
=I/\lambda _{J}$, where $\lambda _{J}$ is the Josephson penetration depth.

In this setting, one is actually dealing with a {\it current dipole}, the
total current traversing the Josephson barrier being zero. The dipole gives
rise to a local magnetic flux $\Phi $ in the region between the injection
and collection points. If the magnitude of this pinned flux attains the flux
quantum $\Phi _{0}$, it may become energetically favorable (in the presence
of an additional dc bias current uniformly distributed along the junction)
to have the pinned flux compensated by a negative flux $-\Phi _{0}$. In
fact, this implies that a fluxon-antifluxon pair is created in the annular
JJ, so that the antifluxon carrying the flux $-\Phi _{0}$ is pinned by the
current dipole, while elsewhere in the long junction there appears a free
fluxon carrying the compensating flux $\Phi _{0}$. Further, if $\Phi $
exceeds $N\Phi _{0}$, then $N$ free fluxons are expected to appear in the
annular JJ.

One objective of this work is to theoretically analyze the above problem and
find a critical condition for the generation of free fluxons by the current
dipole in a long JJ biased by the uniformly distributed dc bias current
$\gamma $. Results will be obtained in an exact analytical form (in section
II) for the case when the dipole strength $\kappa \equiv \varepsilon d$ is
finite, while $d\rightarrow 0$ and $\varepsilon \rightarrow \infty $. Two
different solutions will be obtained. One of them yields a critical value of
$\gamma $ which coincides with the Fraunhofer expression (see below), that
was recently found as a condition for the transition into a resistive state
in a short annular JJ, whose full length was assumed to be much smaller than
$\lambda _{J}$ \cite{Nappi2002}. In our solution, the length of the ring
junction may be arbitrarily large. Thus, we demonstrate that the Fraunhofer
formula for the critical bias current is a {\em universal} one, which is
valid irrespective of the dipole strength and junction's length.

The second analytical solution reported in this paper is exact only for long
junctions. It yields the critical value of $\gamma $ which is slightly
smaller than the one given by the first solution. However, the configuration
on which the second solution is based (essentially, it is a
fluxon-antifluxon bound state) is, plausibly, always unstable. For this
reason, the second solution may be irrelevant to the experimental technique
outlined above in the usual (classical) regime of operation. However,
precisely this second solution defines an {\it instanton} \cite{instanton}
which controls the tunneling rate for quantum birth of a free fluxon, in the
same long JJ, at ultra-low temperatures. The instanton predicts that the
exponentially small factor, which limits the quantum-birth rate, will scale
as a square root of that for the long uniform junction, i.e., the quantum
yield is strongly enhanced by the current dipole. Therefore, the second
solution is physically meaningful too. The analytical predictions will be
compared with results of direct numerical simulations of the critical
condition (in the classical situation).

A free fluxon moving in the annular JJ will periodically collide with the
antifluxon pinned by the current dipole and the dipole itself. An important
issue is to minimize interaction between the free fluxon and the
dipole-antifluxon complex, so that the junction would seem as uniform as
possible for the moving soliton. In section III, we demonstrate, in an
approximate analytical form, that, keeping the dipole size $d$ and current
$\varepsilon $ finite (or proceeding to the limit of the point-wise dipole),
one may indeed select their values so that the interaction is strongly
suppressed. This prediction, which has immediate relevance to the
experiment, is also compared with direct numerical simulations.

\section{Critical conditions for generation of free classical and quantum
fluxons}

\subsection{General considerations}

The system outlined above is described by the following perturbed
sine-Gordon (sG) equation for the superconducting phase difference $\phi $,
which is written in the standard notation \cite{Barone,Ust-rev98}:
\begin{eqnarray}
&&\phi _{tt}-\phi _{xx}+\sin \phi =  \nonumber \\
&&-\alpha \phi _{t}-\gamma -\varepsilon \left[ \delta \left( x-d/2\right)
-\delta \left( x+d/2\right) \right] ,  \label{sG}
\end{eqnarray}
where $\delta $ is the Dirac delta-function [it is used in Eq. (\ref{sG})
due to the assumption, which corresponds to the experimental situation, that
the injection and collection leads have their width much smaller than
$\lambda _{J}$], $\gamma$ is the density of the uniformly distributed dc bias
current, and $\alpha $ is a dissipative constant. In the limit of the
point-like dipole ($d\rightarrow \,0$), the limiting form of Eq.~(\ref{sG})
is
\begin{equation}
\phi _{tt}-\phi _{xx}+\sin \phi =-\alpha \phi _{t}-\gamma +\kappa \delta
^{\prime }(x),  \label{dipole}
\end{equation}
where $\kappa \equiv \varepsilon \,d$ is the dipole strength defined above,
and $\delta ^{\prime }$ is the derivative of the delta-function.

The model (\ref{dipole}) has been first studied almost two decades ago by
Aslamazov and Gurovich \cite{Gurovich-JETPLett84}. These authors considered
interaction of fluxons with an Abrikosov vortex trapped in one of the
junction's electrodes, so that the vortex' normal core is parallel to the
tunnel barrier. Later, the influence of the Abrikosov vortex on the fluxon
bound states localized near the junction boundary was studied by Fistul and
Giuliani \cite{Fistul-Guliani98a}.

In the special case $\kappa =\pi $, Eq.~(\ref{dipole}) is also used as a
model which describes composite long JJs including segments of $0$-type and
$\pi $-type Josephson barriers \cite{Edik_pi}. Such composite junctions can
be fabricated, for example, by using zigzag barriers between $s$-wave and
$d$-wave superconductors \cite{Smilde:ZigzagPRL}. This model gives rise to
stable semi-fluxons ($\pi $-fluxons with a semi-integer topological charge).
As it was pointed out by Goldobin {\it et al}. \cite{Edik_pi}, semi-fluxons
can also be created in a conventional long junction with local current
injection in the form of a current dipole.

The critical condition for the transition to a free moving fluxon is defined
in the following way: it is necessary to find a maximum value $\gamma _{\max
}$ of the bias current density $\gamma $ such that, for given $\kappa $,
static (time-independent) solutions to Eq.~(\ref{dipole}) exist for $\left|
\gamma \right| <\gamma _{\max }$, and do not exist if $\left| \gamma \right|
$ exceeds $\gamma _{\max }$. As it was clearly shown by dint of direct
simulations of Eq.~(\ref{sG}) reported in Ref.~\onlinecite{AVU}, the
disappearance of the static solution means the appearance of freely moving
fluxon(s) in the long JJ.

Double integration of the static version ($\phi _{t}=0$) of
Eq.~(\ref{dipole}) in a vicinity of the point $x=0$ yields the
following boundary conditions
(b.c.) to be satisfied at this point:
\begin{equation}
\phi (x=+0)-\phi (x=-0)=\kappa ;\,\,\phi ^{\prime }(x=+0)=\phi ^{\prime
}(x=-0)\,.  \label{b.c.}
\end{equation}
Off the point $x=0$, the static solution obeys the equation
\begin{equation}
-\phi ^{\prime \prime }+\sin \phi +\gamma =0.  \label{static}
\end{equation}
Further analysis of the static problem based on Eqs. (\ref{b.c.}) and (\ref
{static}) will be presented in the next subsection.

In the sG system operating in the quantum regime (i.e., in a long JJ kept at
extremely low temperatures), fluxon-antifluxon pairs can be produced as a
result of under-barrier tunneling \cite{Maki}. The corresponding tunneling
rate is determined by an instanton solution, which starts with some static
field configuration (which is just a flat phase distribution, in the case of
a homogenous system) and, going in imaginary time under the energy barrier,
ends up on the physical shell, with a state consisting of far separated
fluxon and antifluxon. In was shown that, in the presence of local
inhomogeneities, the quantum birth of fluxon-antifluxon pairs may be
facilitated (the corresponding tunneling rate being enhanced by an
exponentially large factor) if one or both solitons appear on the {\it mass
shell} (appear as real quasi-particles after completion of the tunneling) in
a state pinned by an inhomogeneity \cite{KrivoyRogers}. In this connection,
it is important to find a threshold $\gamma _{{\rm thr}}$ (a minimum value
of $\gamma $) past which there appears a state with a pinned fluxon and free
antifluxon, i.e., an effective threshold for the quantum birth of fluxons in
the quantum regime. A difference from the problem of finding the
above-mentioned value $\gamma _{\max }$, which determines the threshold for
the creation of a free fluxon in the classical regime, is that, on the way
from $\gamma =0$ to $\gamma =$ $\gamma _{\max }$, the corresponding static
solutions to Eq. (\ref{dipole}) need not be stable; just on the contrary,
instanton solutions usually go through unstable states, such as
fluxon-antifluxon pairs, as the evolution in imaginary time does not require
stability in real time \cite{Maki,KrivoyRogers}.

\subsection{Critical current in the annular junction of arbitrary length,
the classical regime}

In the case of a finite-length annular JJ, the above-mentioned static
problem takes the following form: one should find a solution to Eq. (\ref
{static}) such that it satisfies b.c. which is tantamount to Eq.
(\ref{b.c.}):
\begin{equation}
\phi (x=L)-\phi (x=0)=\kappa ;\,\,\phi ^{\prime }(x=L)=\phi ^{\prime
}(x=0)\,,  \label{L}
\end{equation}
where $L$ is the full length of the annular junction. Irrespective of the
length, Eq. (\ref{dipole}) is equivalent to the Newton's equation of motion
in ``time'' $x$ for a particle with the coordinate $\phi $, in the presence
of the potential
\begin{equation}
U(\phi )=-\gamma \phi +\cos \phi ,  \label{U}
\end{equation}
hence the motion conserves the Hamiltonian,
\begin{equation}
H=\frac{1}{2}\left( \frac{d\phi }{dx}\right) ^{2}+\cos \phi +\gamma \phi .
\label{H}
\end{equation}
In the annular junction of any length, the values of $H$ on both sides of
the matching point, $x=0$, must be equal, as they belong to one and the same
solution. Besides that, due to the continuity of $d\phi /dx$ at the matching
point $x=0$, see Eqs. (\ref{b.c.}), the values of $\left( d\phi /dx\right)
^{2}$ are also equal on both sides. Thus, equating the values of the
Hamiltonian, we obtain a condition
\begin{equation}
\cos \left( \phi _{1}+\kappa \right) +\gamma \left( \phi _{1}+\kappa \right)
=\cos \phi _{1}+\gamma \phi _{1}\,,  \label{matching}
\end{equation}
where $\phi _{1}$ is the value on $\phi $ on one left side of the matching
point, the value on the right side being $\phi _{2}=\phi _{1}+\kappa $,
according to Eqs. (\ref{b.c.}).

A more convenient form of Eq. (\ref{matching}) is
\begin{equation}
\gamma =\kappa ^{-1}\left[ \left( 1-\cos \kappa \right) \cos \phi _{1}+(\sin
\kappa )\sin \phi _{1}\right] \,.  \label{sincos}
\end{equation}
One can now look for the largest possible value of the expression on the
right-hand side of Eq. (\ref{sincos}) that can be obtained by varying $\phi
_{1}$. This yields an expression of the Fraunhofer's type for the critical
bias-current density,
\begin{equation}
\gamma _{{\rm c}}=2\kappa ^{-1}\sin \left( \kappa /2\right) ,  \label{Frau}
\end{equation}
which is attained at $\phi _{1}=\pi /2-\kappa /2$.

Thus, Eq. (\ref{Frau}) gives the largest value of $\gamma $ beyond which no
static solution may exist in a vicinity of the point dipole in the annular
JJ of any length, implying a transition to a dynamical regime, i.e.,
generation of free fluxon(s) in the case of the long JJ. Note, however, that
it may happen that more than one critical values exist, then Eq.
(\ref{Frau}) gives, according to its derivation from Eq. (\ref{sincos}),
only the {\em
largest} one among them. In fact, it will be shown in the next subsection
that (in the case of the long JJ) there indeed exists exactly one more
critical value, which is smaller than (\ref{Frau}), see Eq. (\ref{solution})
below.

Quite naturally, in the limit $\kappa \rightarrow 0$ Eq. (\ref{Frau}) yields
$\gamma _{{\rm c}}=1$, which is the commonly known critical value of the
bias current density in the long homogeneous JJ \cite{Barone}.The fact that
$\gamma _{{\rm c}}$ vanishes at $\kappa =2\pi $ is also easy to understand:
the static equation (\ref{static}) has obvious stable uniform solutions,
\begin{equation}
\phi _{0}^{(n)}=-\sin ^{-1}\gamma +2\pi n  \label{phi0}
\end{equation}
(provided that $|\gamma |\leq 1$), with an arbitrary integer $n$, hence the
b.c. (\ref{b.c.}) with $\kappa =2\pi $ implies that two such solutions,
$\phi _{0}^{(0)}$ and $\phi _{0}^{(1)}$, match to each other across the point
$x=0$; in the annular system, they must be connected by a static
$2\pi$-kink, which is possible exactly at $\gamma =0$. Similarly, in the case
$\kappa =2\pi N$ with any integer $N$, Eq. (\ref{Frau}) again yields $\gamma
_{{\rm c}}=0$, which means that $N$ fluxons will appear spontaneously in
this case, without the application the bias current.

The result (\ref{Frau}) {\em does not} depend on the length of the annular
junction; for this reason, it coincides with the threshold for transition to
a resistive state in a {\em short} annular JJ, which was found in the recent
work \cite{Nappi2002}. It is relevant to mention that the actual calculation
of the threshold in that work was performed in an altogether different way,
so that the sG equation was not used at all.

\subsection{The fluxon-birth threshold in the long annular junction, the
quantum regime}

Coming back to the case of the long JJ, we notes that a solution for $\phi
(x)$ must assume the background value (\ref{phi0}) (for instance, with
$n=0$) far from the dipole (formally, at $x\rightarrow \pm \infty $). Thus, in
the case of a very long (formally, infinite) junction, we need to find two
solutions of Eq.~(\ref{static}), in the regions, respectively, $x>0$ and $x<0
$, so that one solution assumes the asymptotic value (\ref{phi0}) at
$x\rightarrow +\infty $, and the other one assumes the same value at
$x\rightarrow -\infty $. The solutions must be matched at the point $x=0$,
pursuant to b.c. (\ref{b.c.}), and it is necessary to find a maximum value
of $\left| \gamma \right| $ for which the matching is possible with a given
dipole strength $\kappa $.

Consideration of the equivalent mechanical problem mentioned above, i.e.,
motion in the potential (\ref{U}), demonstrates that there are {\em exactly
two} solutions to this problem. One of them yields the above result (\ref
{Frau}). A description of both solutions (for the case of a very long JJ) is
given below.

Both solutions employ the single function $\phi (x)$ which satisfies Eq.
(\ref{static}) and assumes the same asymptotic value (\ref{phi0})
simultaneously at $x\rightarrow +\infty $ and $x\rightarrow -\infty $. This
function represents a fluxon-antifluxon (${\rm f}\overline{{\rm f}}$) bound
state; it is an even function, with a maximum value $\phi _{\max }$ at the
central point. Using the conservation of the Hamiltonian (\ref{H}), one can
find a relation between $\phi _{\max }$ and $\phi _{0}$, which takes the
form
\begin{equation}
\gamma \sin ^{-1}\gamma +\cos \left( \sin ^{-1}\gamma \right) =-\gamma \phi
_{\max }+\cos \phi _{\max }\,.  \label{phimax}
\end{equation}

In addition to the ${\rm f}\overline{{\rm f}}$ solution, Eq. (\ref{static})
also has a semi-divergent one, which starts with the value (\ref{phi0}),
say, at $x=-\infty $, and diverges as $\phi (x)\approx \left( \gamma
/2\right) x^{2}$ at $x\rightarrow +\infty $. This solution may also be
employed to construct a full static state consisting of two pieces that are
matched as per Eqs. (\ref{b.c.}) at $x=0$. It is easy to check that the
state built as the combination of the ${\rm f}\overline{{\rm f}}$ and
semi-divergent solutions disappears, with the increase of $\gamma $, exactly
at the critical point (\ref{Frau}), so this is how the solution
corresponding to the Fraunhofer's formula (\ref{Frau}) looks in the long JJ.

Besides this matched state, there is another one, specific to the long
junction. It  can be constructed from two pieces belonging to the
${\rm f}\overline{{\rm f}}$ solution [it is easy to see that no state satisfying
the b.c. (\ref{b.c.}) can be constructed if both pieces belong to the
semi-divergent solution; thus no extra solution is possible in addition to
the two presently considered ones].

\begin{figure}[tbh]
\centering
\includegraphics[width=3.0in]{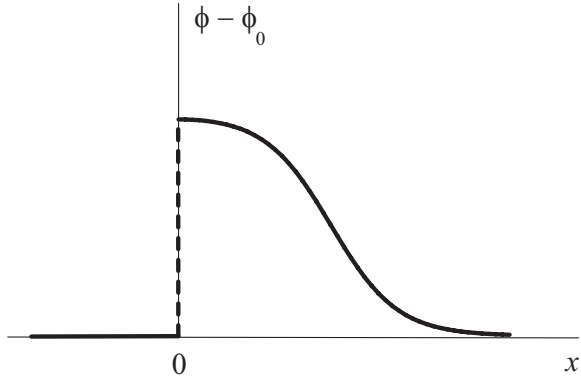}
\caption{A sketch of the critical field configuration that gives rise to
the threshold value (\ref{solution}) of the bias-current density.}
\label{extrafig}
\end{figure}

At the critical point corresponding to the disappearance of the new matched
state, the b.c. (\ref{b.c.}) are satisfied in an extreme form: on one side
of the point $x=0$, we have the uniform background solution, $\phi \equiv
\phi _{0}$, while on the other side a piece of the ${\rm f}\overline{{\rm f}}$
solution is to be used. Due to the continuity of the derivative $\phi
^{\prime }(x)$ at $x=0$ [see Eqs.~(\ref{b.c.})] and the fact that $\phi
^{\prime }$ vanishes at the matching point in the critical case [as the
derivative of $\phi (x)\equiv \phi _{0}$ is zero], the central point (the
one with $\phi =\phi _{\max }$) of the ${\rm f}\overline{{\rm f}}$ solution,
which is the only spot where the derivative vanishes, must be set exactly at
$x=0$ when one attains the critical configuration, see
Fig. \ref{extrafig}. In view of the first b.c.
(\ref{b.c.}), this means that the background value (\ref{phi0}) and $\phi
_{\max }$ are related (in the critical state) so that $\phi _{\max }=\kappa
+\phi _{0}\equiv \kappa -\sin ^{-1}\gamma $. This expression should be
inserted in Eq.~(\ref{phimax}). After that, the ``worst'' transcendental
terms $\gamma \sin ^{-1}\gamma $ in the ensuing equation mutually cancel,
and the remaining equation can be solved in an exact form, to yield a new
critical value of $\gamma $, which, this time, we call a threshold one:
\begin{equation}
\gamma _{{\rm thr}}=\frac{2\sin ^{2}(\kappa /2)}{\sqrt{4\sin ^{4}(\kappa
/2)+\left( \kappa -\sin \kappa \right) ^{2}}}\,.  \label{solution}
\end{equation}

As well as in the case of the Fraunhofer expression (\ref{Frau}), the limit
value of $\gamma _{{\rm thr}}$ for $\kappa \rightarrow 0$ is $1$, which is
the above-mentioned critical bias-current density in the uniform long JJ. On
the other hand, if $\kappa $ is close to $2\pi $, i.e., the phase jump in
b.c. (\ref{b.c.}) is close to $2\pi $, the ${\rm f}\overline{{\rm f}}$
solution takes the form of a bound state of a fluxon and an antifluxon
separated by a large distance, which implies that even a small bias current
is able to destroy this configuration pinned by the dipole. Accordingly, one
finds from Eq.~(\ref{solution}) that
\begin{equation}
\gamma _{{\rm thr}}\approx \left( 4\pi \right) ^{-1}\left( \kappa -2\pi
\right) ^{2}  \label{minimum}
\end{equation}
when $\kappa $ is close to $2\pi $, which should be compared to the
asymptotic form of the Fraunhofer critical value in the same case: as it
follows from Eq. (\ref{Frau}), $\left| \gamma _{{\rm c}}\right| \approx
\left| \kappa -2\pi \right| /8$. On the contrary to this linearly vanishing
expression, Eq.~(\ref{minimum}) shows that the newly found threshold value
vanishes quadratically. As well as the Fraunhofer critical value, the
threshold one (\ref{minimum}) has higher-order minima at $\kappa =2\pi N$
with $N$ integer.

The comparison of the two critical values (versus $\kappa $) is shown below
as a part of Fig. \ref{num2}(a). As is seen from the figure (and can be
proved analytically), the threshold value (\ref{minimum}) is always smaller
(except for the points $\kappa =2\pi N$, where both critical values vanish),
which implies that, in the application to the tunnel-birth problem for
fluxons in the quantum system, it will be a dominating one. In principle,
for the same reason it could play a dominant role in the transition to the
dynamical regime in the classical case, but, in fact, this is hindered by
the instability of the ${\rm f}\overline{{\rm f}}$ static configuration
which underlies Eq. (\ref{solution}). Unlike this, it was checked in direct
simulations that the configuration based on the matched ${\rm f}\overline{
{\rm f}}$ and semi-divergent solutions, which yields the Fraunhofer's
expression (\ref{Frau}), is always stable in the long junction (for the case
$\kappa =\pi $, it was also shown in Ref. \cite{Edik}). However, as it was
explained above, the dynamical instability of the intermediate state does
not bar using the configuration as a part of the under-barrier instanton
trajectory.

While detailed consideration of the under-barrier production of
fluxons in the quantum system is beyond the scope of this work, we
briefly mention that the above-mentioned instanton may be
constructed, following the pattern of the critical configuration
that yields the threshold value (\ref{solution}), as precisely a
{\em half} of the fluxon-antifluxon bound state on one side of the
point $x=0$, matched to the uniform state (\ref{phi0}) on its
other side. This makes the integral of the system's action, taken
along the under-barrier evolution in the imaginary time, which
determines the exponential smallness of the tunneling rate
\cite{Maki}, equal to exactly half the integral in the absence of
the current dipole. For this reason, the exponential smallness is
expected to scale as the {\em square root} of that in the uniform
system, so that the tunneling will be strongly enhanced. As for
the direct role of the threshold value (\ref{solution}), it
implies that, if $\gamma $ is taken larger than it, the tunneling
mechanism will directly generate a free fluxon in the system.

\section{Suppression of the interaction between a free fluxon and the
antifluxon pinned by the current dipole}

As it was mentioned in the Introduction, another problem relevant to the
experiment is to suppress interaction between an already existing free
fluxon and its antifluxon counterpart, which is pinned by the dipole pair of
currents in the long JJ. A natural way to select parameters of the dipole
configuration in Eq.~(\ref{sG}) with, generally speaking, {\em finite} $d$
and $\varepsilon $, so that to minimize the interaction, is to consider the
case when the distance between the free fluxon and pinned antifluxon is
large enough, so that the interaction between them may be treated by means
of the perturbation theory \cite{review}. In the framework of this approach,
the free fluxon with zero velocity, whose form is
\begin{equation}
\phi _{{\rm fl}}=4\tan ^{-1}\left[ \exp \left( \sigma (x-\xi \right) \right]
\,,  \label{fluxon}
\end{equation}
where $\xi $ is the coordinate of the fluxon's center, and $\sigma =\pm 1$
is its polarity, is regarded as a quasi-particle. Since the term in the
Hamiltonian of the full sG model (\ref{sG}), which corresponds to the
current-dipole terms in the equation, is $H_{{\rm dip}}=\varepsilon \left[
\phi (x=d/2)-\phi (x=-d/2)\right] $, it is easy to obtain an effective
potential of direct interaction of the free fluxon with the current dipole,
substituting the waveform (\ref{fluxon}) in $H_{{\rm dip}}$. Eventually, one
can find an effective force of the direct fluxon-dipole interaction,
\begin{eqnarray}
F_{{\rm dip}} &\equiv &-\frac{dH_{{\rm dip}}}{d\xi }=2\sigma \varepsilon
\left[ {\rm sech}\left( \xi -d/2\right) -{\rm sech}\left( \xi +d/2\right)
\right]   \nonumber \\
&\approx &8\sigma \varepsilon \,\sinh (d/2)\,\,e^{-\xi },  \label{direct}
\end{eqnarray}
where we made use of the assumption that the distance $\left| \xi \right|$
of the fluxon from the dipole is large. The polarity $\sigma $ of the free
fluxon and the sign of the current $\varepsilon $ mist correlate so that the
free fluxon is repelled by the dipole, while the corresponding antifluxon is
attracted by the it (otherwise, the antifluxon cannot be in the pinned
state). As it follows from Eq. (\ref{direct}), this implies that $\sigma =
{\rm sign}\,\varepsilon $.

Besides that, the fluxon also interacts with the antifluxon pinned by the
dipole. A well-known perturbative expression for the interaction force
between the fluxon and antifluxon is \cite{review} $F_{{\rm antifl}}\approx
-32\,e^{-\xi }$. Then, a simple condition for the effective suppression of
the net interaction between the free fluxon and the complex including the
current dipole and the antifluxon pinned by it may be formulated as a
condition for\ mutual cancellation of the two forces, the repulsive one
$F_{{\rm dip}}$ and the attraction force $F_{{\rm antifl}}$, which produces the
following result:
\begin{equation}
\left| \varepsilon \right| \,{\rm \sinh }\left( d/2\right) =4.
\label{result}
\end{equation}
The meaning of this result is that the free fluxon is expected to move
nearly as in the homogeneous long JJ if the parameters of the current dipole
are selected according to Eq.~(\ref{result}). Note also that, in the limit
of the point-like dipole, which was dealt with in the previous section,
i.e., $\varepsilon \rightarrow \infty $ and $d\rightarrow 0$, the condition
(\ref{result}) takes a very simple form, $\kappa =8$. This result is obtained
in the framework of the approximations adopted above, but it is not very
different from the exact result that follows from Eq.~(\ref{Frau}), i.e.,
$\kappa =2\pi $ (at this value of $\kappa $, the fluxon can be very easily
separated from the dipole-antifluxon complex).

The validity of the prediction (\ref{result}) was checked against direct
numerical simulations of the full equation (\ref{sG}). For instance, in the
case shown below in Fig. \ref{3D-num}, with $\varepsilon =8$ and $d=0.5$,
the product on the left-hand side of Eq. (\ref{result}) takes the value $4.17
$, and the interaction with the dipole-antifluxon complex indeed gives rise
to a very small perturbation in motion of the free fluxon. Simulation run
for values of $\left| \varepsilon \right| \,{\rm \sinh }\left( d/2\right) $
quite different from $4$ (not shown here, as the pictures are rather messy)
demonstrate a much stronger perturbation.

\section{Numerical simulations}

In order to verify analytical results for the fluxon injection, we performed
numerical simulations by solving the full equation (\ref{sG}). In the
simulations, each $\delta $-function in Eq.~(\ref{sG}) was approximated by
its smooth counterpart,

\begin{equation}
\varepsilon \delta (x)\approx \eta g(x)\equiv \eta \,{\rm sech}^{2}\left(
2x/\xi \right),  \label{appr-delta}
\end{equation}
such that $\eta \xi =\varepsilon $, which complies with the definition
$\int_{-\infty }^{+\infty }\delta (x)dx=1$. This approximation implies that
the injected current is spread over the distance $\simeq \xi \lambda _{{\rm
J }}$. We will present results obtained with the values $\xi =1$ and $\xi
=0.1$ , which lie in a typical experimentally accessible range of this
parameter \cite{AVU}. The numerically calculated current-voltage
characteristics will be shown in normalized units as $\gamma (v)$, were $v$
is the average fluxon velocity normalized to the Swihart velocity $\bar{c}$.
With this normalization $v=1$ corresponds to the asymptotic voltage of the
single-fluxon step. The simulations are performed with the normalized
junction length $L=10$ and dissipation coefficient $\alpha =0.1$.

\begin{figure}[tbh]
\centering
\includegraphics[width=3.3in]{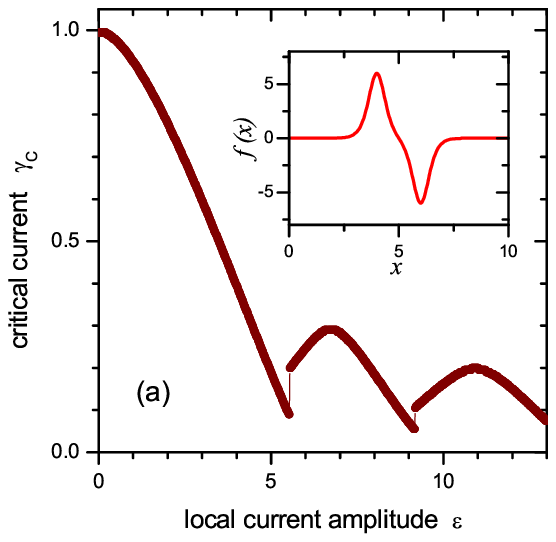}
\centering
\includegraphics[width=3.3in]{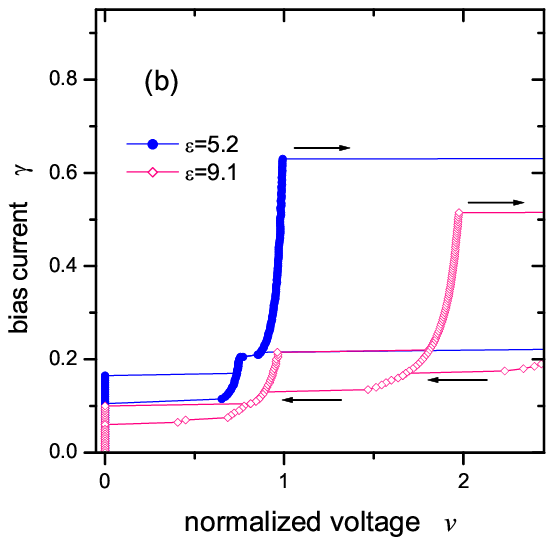}
\caption{(a) Numerically calculated dependence of the critical
current of the annular junction $\protect\gamma_{{\rm c}}$ on the
injection-current amplitude $\protect\varepsilon$ for $d=2$. (b)
Numerically calculated current-voltage characteristics of the
annular junction for two values of the injection current,
$\protect\varepsilon=5.2$ and $\protect\varepsilon =9.1 $.}
\label{num1}
\end{figure}

Figure~\ref{num1}(a) presents the calculated dependence of the critical
current $\gamma _{{\rm c}}$ on the injection-current amplitude $\varepsilon$
for $d=2$. The inset shows the adopted approximation for the
injected-current profile, $f(x)=\eta \left[ g\left( x-d/2\right) -g\left(
x+d/2\right) \right] $, see Eq. (\ref{appr-delta}), with $\xi =1$ and $\eta
=9$. As it has already been concluded from experimental data \cite{AVU}, the
dependence in Fig.~\ref{num1} is very similar to the conventional Fraunhofer
pattern of the critical current in a small Josephson junction in the
magnetic field, the length of the equivalent small junction being associated
with the distance $d$ between the injecting points. The overlap between the
lobes gets larger for larger $d$. Due to this overlap, the minimum value of
the critical (fluxon-depinning) current between the lobes decreases with $d$.

Numerically calculated $I$-$V$ curves for various values of the injection
current (indicated in the plots) are shown in Fig.~\ref{num1}(b). One-fluxon
and two-fluxon steps can be clearly recognized here. Numerical data show
that the steps on $I$-$V$ curves are accounted for by fluxons freely moving
under the action of the uniformly distributed bias current. It can be noted
from Fig.~\ref{num1}(a) that there is a residual pinning of fluxon(s) due to
the disturbance produced by the current injectors. This pinning is smallest
at injection-current values which lie between the lobes of the $\gamma_{
{\rm c}}(\varepsilon )$ curve. Thus, for a given injector spacing $d$, the
residual fluxon pinning can be minimized by choosing an appropriate value
for the injection current $\varepsilon $.

\begin{figure}[tbh]
\centering
\includegraphics[width=3.3in]{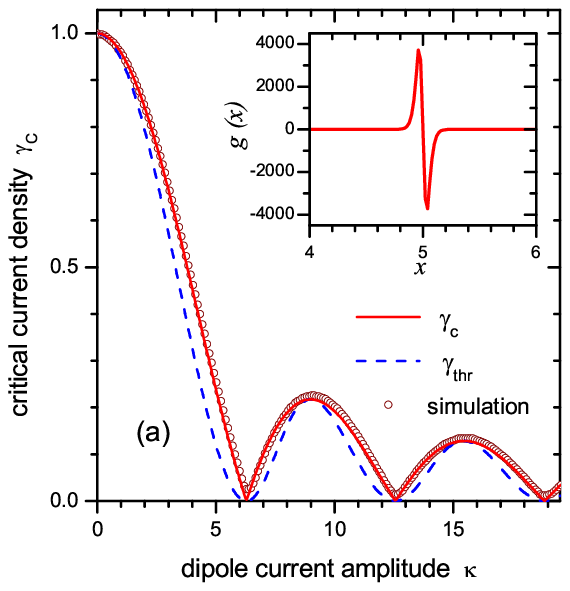}
\centering
\includegraphics[width=3.3in]{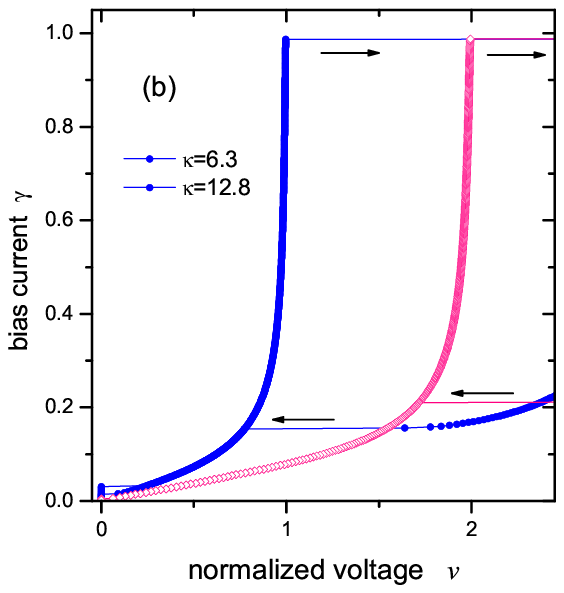}
\caption{(a) Numerically calculated dependence of the critical
current in the annular junction, $\protect\gamma_{{\rm c}}$, on
the dipole-current amplitude $\protect\kappa$. The newly found
threshold value, given by Eq.~(\ref{solution} (dashed line), and
the Fraunhofer value, given by Eq. (\ref{Frau}) (solid line) are
included too. (b) Numerically calculated current-voltage
characteristics of the annular junction for two values of the
dipole's strength, $\protect\kappa=6.3$ and
$\protect\kappa=12.8$.} \label{num2}
\end{figure}

In order to numerically model the point-like dipole given by the last term
of Eq.~(\ref{dipole}), we used the derivative $g^{\prime }(x)$ of the
function $g(x)$ defined by Eq.~(\ref{appr-delta}). Figure~\ref{num2}(a)
shows the calculated dependence $\gamma _{{\rm c}}$ on the dipole-current
amplitude $\kappa $. The inset shows the numerically used current profile,
$f(x)\equiv \eta g^{\prime }\left( x-d/2\right) $ with $\xi =0.1$. The
spatial grid used in this calculation had $\Delta x=0.01$. In spite of the
limited accuracy of our numerical scheme because of the sharp current
profile, the agreement between the numerical data and the Fraunhofer formula
(\ref{Frau}), represented by the solid line, is very good. For the
comparison's sake, in Fig.~\ref{num2}(a) the dashed line shows the
newly-found threshold value given by Eq.~(\ref{solution}).

The numerically calculated $I$-$V$curves for the point-like dipole are shown
in Fig.~\ref{num2}(b). In comparison with Fig.~\ref{num1}(b), the depinning
current is very small, and the value of the bias-current density
corresponding to the steps is very close to the maximum value, $\gamma =1$.
Both the one-fluxon and two-fluxon steps can be hardly distinguished from
those in the ideal uniform annular junction with trapped fluxons.

\begin{figure}[tbh]
\centering
\includegraphics[width=3.3in]{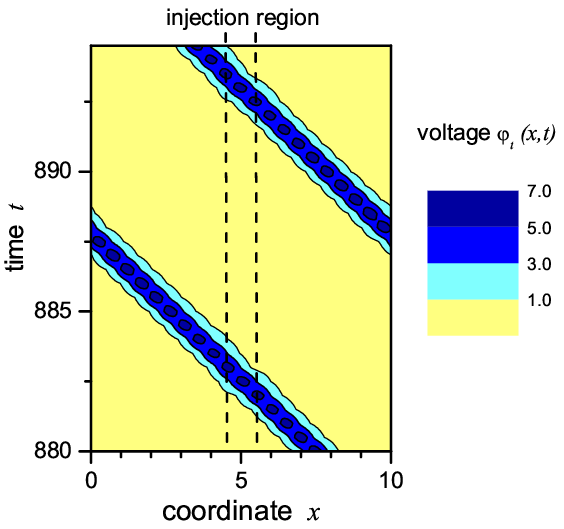}
\caption{Spatially-temporal evolution of the normalized
instantaneous voltage in simulations of the annular Josephson
junction with $\varepsilon=8$, $d=1$ and $\protect\gamma=0.4$.
Dashed lines indicate the current injection points.}
\label{3D-num}
\end{figure}

The presentation of numerical results is concluded by Fig.~\ref{3D-num}. It
shows a 2D gray scale-plot of the spatially-temporal evolution of the
instantaneous normalized voltage $\varphi _{t}(x,t)$ in the annular junction
for $\varepsilon =8$, $d=1$, and $\gamma =0.4$. The moving fluxon is
recognized as a solitary-wave packet propagating at a nearly constant
velocity across the junction. One can see that the disturbance of the fluxon
motion in the region where the current injector is located is tiny, which
was explained above, by the proximity of this case to the
interaction-suppression condition (\ref{result}). The residual perturbation
may, nevertheless, be significant at small velocities of the fluxon, when
its kinetic energy is comparable to the pinning potential generated by the
perturbation.

\section{Conclusion}

In this work, we have considered static and dynamical solutions
for fluxons in the model of the annular Josephson junction with
the local current dipole. The analysis is relevant for
interpreting the recent experiment \cite{AVU} that implemented a
new technique of inserting fluxons into long junctions.

Recently, a critical value of the bias-current density $\gamma $,
above which the system performs transition to a dynamical state,
was found \cite{Nappi2002}, in the Fraunhofer's form, for annular
junctions whose length is much smaller than $\lambda _{J}$. We
have shown that, in the limit of the infinitely narrow current
dipole, the same critical expression is valid{\em \ at all values
}of the junction's length. In the long junction, the dipole
generates free fluxon(s) when $\gamma $ exceeds the critical value
$\gamma _{ {\rm c}}$. We have also found another critical value
$\gamma _{{\rm thr}}$ for long junctions, which is always slightly
smaller than the Fraunhofer's one, except for points where both
vanish. The phase configuration which yields the new critical
value is generated by an unstable fluxon-antifluxon bound state,
therefore it is not relevant to the classical-fluxon-generation
regime. However, it determines a quantum-birth threshold for
fluxons in the quantum regime. It was also concluded that these
two critical values exhaust all possible solutions to the problem
of finding the critical bias-current density in the presence of
the current dipole.

The interaction of a free fluxon with the complex consisting of the current
dipole and antifluxon pinned by it was considered too. A condition for
suppression of the effective interaction force was predicted, which makes
the long junction effectively uniform for the motion of the free fluxon.

The analytical predictions obtained in this work were checked against direct
simulations. In all the cases, they agree well.

\section{Acknowledgements}

We appreciate valuable discussions with E. Goldobin. B.A.M. acknowledges
financial support from DAAD (Deutscher Akademischer Austaschdienst) and
hospitality of the Physics Institute at the Universit\"{a}t
Erlangen-N\"{u}rnberg.

\end{document}